\documentclass[twocolumn,prb,floatfix,superscriptaddress]{revtex4-2}
\usepackage{amsmath}
\usepackage{amssymb}
\usepackage{graphicx}
\usepackage{bm}
\usepackage{color}
\usepackage{hyperref}
\usepackage{upgreek}
\usepackage{scalerel}
\usepackage{pgffor}
\usepackage{pdfpages}
\usepackage{float}
\usepackage{physics} 
\usepackage{lscape}   
\usepackage{comment}  
\makeatletter
\AtBeginDocument{\let\LS@rot\@undefined}
\makeatother

\begin{document}
\title{Disorder-robust $p$-wave pairing with odd frequency dependence in normal metal-conventional superconductor junctions}
\author{Tomas L\"{o}thman}
\affiliation{Department of Physics and Astronomy, Uppsala University, Box 516, S-751 20 Uppsala, Sweden}
 \author{Christopher Triola} 
 \affiliation{Los Alamos National Laboratory, Los Alamos, New Mexico 87544, USA}
  \author{Jorge Cayao} 
 \email[Corresponding author: ]{jorge.cayao@physics.uu.se}
\affiliation{Department of Physics and Astronomy, Uppsala University, Box 516, S-751 20 Uppsala, Sweden}
 \author{Annica M. Black-Schaffer}
\affiliation{Department of Physics and Astronomy, Uppsala University, Box 516, S-751 20 Uppsala, Sweden}

\date{\today} 
\begin{abstract}
We investigate the induced superconducting pair correlations in junctions between a conventional spin-singlet $s$-wave superconductor and a disordered normal metal. Decomposing the pair amplitude based on its symmetries in the time domain, we demonstrate that the odd-time, or equivalently odd-frequency, spin-singlet $p$-wave correlations are both significant in size and entirely robust against random non-magnetic disorder. We  find that these odd-frequency correlations can even be generated by disorder.  Our results show that anisotropic odd-frequency pairing  represent an important fraction of the proximity-induced correlations in disordered superconducting hybrid structures. 
\end{abstract}\maketitle

%%%%%%%%%%%%%%%%%%%%%%%%%%%%%%%%%%%
%                                         INTRODUCTION                                      %
%%%%%%%%%%%%%%%%%%%%%%%%%%%%%%%%%%%
\section{Introduction}%
\label{sec1}
Superconductivity arises from the formation and condensation of a macroscopic number of electron pairs, or Cooper pairs. 
The superconducting state is characterized by the Cooper pair amplitude which, due to the fermionic nature of electrons, is antisymmetric under the exchange of all degrees of freedom describing the paired electron states. These symmetries include the positions, spin, and relative time separation of the two electrons.   

In BCS theory of superconductivity the interparticle interaction is instantaneous and, hence, only static, equal-time, pair amplitudes contribute to the order parameter. Such equal-time pair symmetries are then constrained to be either even in space and odd in spin, as in the conventional spin-singlet $s$-wave state, or odd in space and even in spin (e.g.~spin-triplet $p$-wave). However, even for BCS superconductors, a growing body of evidence suggests that dynamical pair correlations that are odd in the relative time \cite{bere74,PhysRevLett.66.1533,PhysRevB.45.13125,PhysRevB.46.8393,PhysRevB.47.513} can play a role in the physics of superconducting heterostructures \cite{RevModPhys.77.1321,Balatsky2017,cayao2019odd}. Due to their odd time-dependence, these dynamical pair correlations must be either even in space and even in spin or odd in space and odd in spin. Such exotic odd-time pairs are equivalently also referred to as odd-frequency (odd-$\omega$) pairs \cite{bere74}.

A well-established host of odd-time/odd-$\omega$ pairs are ferromagnet-superconductor (FS) junctions. In these systems, the combination of spatial invariance breaking due to the interface and spin-singlet to spin-triplet conversion due to the ferromagnet, generates finite odd-$\omega$ spin-triplet $s$-wave pairs in the F region even for conventional spin-singlet $s$-wave BCS superconductors \cite{PhysRevLett.86.4096,Kadigrobov01}.  The presence of odd-$\omega$ $s$-wave pairs explains several unconventional features of FS junctions, such as the long-range proximity effect \cite{longrangeExp,PhysRevB.58.R11872,Keizer06,PhysRevLett.104.137002,wang10,Robinson59,PhysRevB.82.100501,Cirillo_2017,PROSHIN2018359}, zero-bias peaks \cite{PhysRevB.92.014508,bernardo15}, and paramagnetic Meissner effect \cite{PhysRevB.64.132507,PhysRevB.64.134506,PhysRevX.5.041021}. 

Odd-$\omega$ pairing can also emerge in normal metal-conventional superconductor (NS) junctions. Here the interface allows for both the original even-$\omega$ spin-singlet $s$-wave (isotropic) and an induced odd-$\omega$ spin-singlet $p$-wave (anisotropic) symmetry in the N region \cite{PhysRevB.76.054522,Eschrig2007,PhysRevB.97.134523,PhysRevB.98.075425,cayao2019odd}. 
However, unlike FS junctions, in   disordered NS systems odd-$\omega$ pairing has so far been largely ignored. 
	One often-cited reason for this neglect is the argument that $p$-wave amplitudes are killed in the presence of disorder \cite{PhysRevLett.98.037003,golubovreview,Nagaosa12}. 
	For example, in the quasiclassical Usadel formulation for dirty systems, only isotropic $s$-wave solutions are explicitly considered, while any anisotropic part is assumed to be negligible \cite{PhysRevLett.25.507}. 
Additionally, Anderson's theorem \cite{ANDERSON195926} has been invoked to defend the assertion of fragility of $p$-wave pairs in disordered NS junctions, even though it does not technically concern proximity-induced effects.

\begin{figure}[!t]
	\centering
	\includegraphics[width=.4\textwidth]{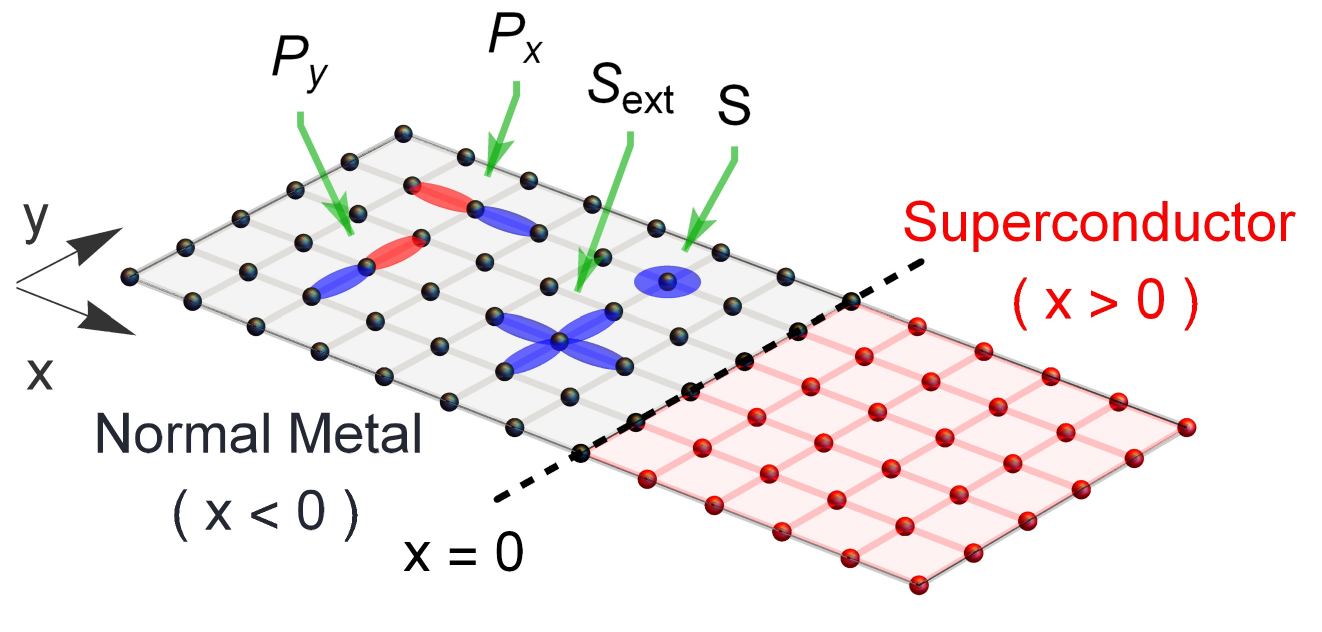}
	\caption{Sketch of a 2D NS junction with interface at $x=0$, where filled blue and red circles represent sites in N ($x<0$) and S ($x>0$) regions. The relevant spatial orbital symmetries, $s$-wave, extended $s$-wave,  and $p_{x,y}$-wave, are depicted in blue (red) for positive (negative) amplitudes.}
	\label{Fig1}
\end{figure}

In this work we explicitly show that in NS junctions odd-$\omega$ spin-singlet anisotropic $p$-wave pair correlations are not only robust against disorder but as disorder strength increases, the odd-$\omega$ amplitudes constitute a growing fraction of the proximity-induced pair correlations in the N region. Remarkably, we even find  odd-$\omega$ pairing generated by disorder.  We do this by considering a NS junction between a disordered N region and a conventional spin-singlet $s$-wave superconductor, see Fig.\,\ref{Fig1}, explicitly calculating the fully quantum mechanical time evolution of the proximity-induced superconducting correlations. 
By showing that odd-$\omega$ $p$-wave pair correlations are ubiquitous and robust, our results provide a completely different view of superconducting pair correlations in disordered heterostructures, with potentially far reaching consequences for the physics, such as a paramagnetic Meissner effect.

The remainder of this paper is organized as follows. In Sec.\,\ref{sect1} we present the model and outline the method to obtain the superconducting pair amplitudes. In Sec.\,\ref{sect2} we present the results for the clean and disordered regimes in 1D and also  show some consequences in 2D. We conclude  in Sec.\,\ref{conclusions}. For completeness, in the Appendices we analytically obtain the induced pair amplitudes in 1D  for the clean regime analytically using perturbation theory. Furthermore, here we also present further numerical calculations  for 2D.

%%%%%%%%%%%%%%%%%%%%%%%%%%%%%%%
%                    MODEL   AND METHODS                              %
%%%%%%%%%%%%%%%%%%%%%%%%%%%%%%%
\section{Model and Methods}
\label{sect1}
To demonstrate the robustness of odd-time/odd-$\omega$ pair correlations we study the proximity-induced pair amplitudes in both one-dimensional (1D) and 2D NS junctions between a disordered N region and a conventional spin-singlet $s$-wave S region, with the interface at $x=0$, see Fig.\,\ref{Fig1}. The length of the N (S) regions are $ L_{\rm N (S)}$, measured in units of the lattice spacing $a=1$. The Hamiltonian for these junctions can be written as
\begin{equation}
\label{Eq:1}
	H=-t\sum_{\langle ij\rangle}c^{\dagger}_{i\sigma}c_{j\sigma}-\mu\sum_{i}c^{\dagger}_{i\sigma}c_{i\sigma}+H_{\Delta}+H_{\rm V},
\end{equation}
where $c_{j\sigma}$ is the destruction operator for a particle with spin $\sigma$ at site $j=(j_{x},j_{y})$ in 2D or $j=j_{x}$ in 1D, $\langle \dots \rangle$ denotes nearest neighbor sites, $t$ the nearest neighbor hopping  amplitude, and $\mu$ the chemical potential  that controls the band filling. The conventional spin-singlet $s$-wave superconductor is captured by $H_{\Delta}=\sum_{i}\theta(x)\Delta c^{\dagger}_{i\uparrow}c^{\dagger}_{i\downarrow}+\text{h.c.}$, with $\Delta$ being the order parameter and $\theta$ the Heaviside step function.  We further introduce random scalar, i.e.~non-magnetic, potential disorder on the normal side of the junction using the Anderson disorder model: $H_{\rm V}=\sum_{i}\theta(-x)V_{i} c^{\dagger}_{i\sigma}c_{i\sigma}$, where the uncorrelated potentials $V_{i}$ are drawn from the uniform box distribution on the interval $[-\delta,\delta]$ and added to every site of our discrete lattice model \cite{PhysRev.109.1492}.
	The disorder strength is controlled by the width of the distribution $\delta/t$ which allows us to tune the disorder from the clean limit or weakly disordered regimes (all sites have none or weak scatterers) to extremely strong disorder (some sites host strong scatterers, while other stays weak), where the disorder dominates over the kinetic energy. 
	This type of disorder can arise from non-magnetic charge impurities or fluctuations in the chemical potential, both inevitable in real experimental samples.

\subsection{Pair amplitudes}
We are interested in the superconducting correlations in NS junctions. These are characterized by the pair amplitudes which can  be obtained from the time-ordered anomalous Green's function $F_{i j \alpha \beta}(t_{2},t_{1})=\langle \mathcal{T} c_{i \alpha}(t_{2}) c_{j \beta}(t_{1}) \rangle$, where $\mathcal{T} $ is the time-ordering operator. The pair amplitude $F_{i j \alpha \beta}(t_{2},t_{1})$ accounts for the propagation of quasiparticles from the ($i \alpha$)-state   at time $t_{1}$ to the ($j \beta$)-state at time $t_{2}$. Since $F_{i j \alpha \beta}(t_{2},t_{1})$  only depends on the relative time difference $\tau=t_{2}-t_{1}$, we focus on the forward time propagation, where the even- (e) and odd-time (o), or equivalently even- and odd-$\omega$, components are extracted as $F^{\rm e(o)}_{i j \alpha \beta}(\tau>0)= (\langle c_{i\alpha}(\tau)c_{j\beta}(0)\rangle \mp \langle c_{j\beta}(\tau)c_{i\alpha}(0) \rangle)/2$. Moreover, with no spin-active fields in NS junctions, the pair amplitudes exhibit the same spin symmetry as the parent superconductor. Hence, the pair amplitudes $F_{i j \alpha \beta}^{\rm e(o)}(\tau)$ can be decomposed into the pair wave amplitudes \cite{Bjoernson2015},
		\begin{equation}
		\label{PAIRX}
			d^{\rm e(o)}_{w} \left(\tau, R \right) = \frac{1}{2} \sum_r  w\left( r \right) 
			\sum_{\alpha \beta} [i\sigma_{y}]^\dagger_{\alpha \beta} F^{\rm e(o)}_{R+r \beta R-r \alpha}(\tau),
		\end{equation}
where we focus our study on the lowest rotation symmetric wave representation projections with $|r| \leq 1$: $w_S(r) = \delta_{|r|=0}$, $w_{S_{ext}}(r) = (1/z)\delta_{|r|=1}$,  $w_{P_{i}} (r) = [(\hat{i} \cdot r)/2] \delta_{|r|=1} $ are the $s$-, extended $s$-wave, and $p_{i}$-wave amplitudes. Here  $z$ is the coordination number, $\sigma_{y}$ the $y$-Pauli matrix, and the center of mass $R=(i+j)/2$, and relative $r=(i-j)/2$ spatial coordinates. %While also higher order representations have a finite magnitude \cite{PhysRevB.76.054522}, our focus on $|r| \leq 1$ allows for a targeted and controlled investigation of the interplay between disorder and anisotropic odd-$\omega$ pair correlations. 
While we expect that higher order representations also have a finite magnitude \cite{PhysRevB.76.054522}, to simplify the presentation we focus on $|r| \leq 1$ which allows for a targeted and controlled investigation of the interplay between disorder and anisotropic odd-$\omega$ pair correlations.

Evaluation of the pair wave amplitudes in each symmetry channel in Eq.~(\ref{PAIRX}) only requires computation of time-dependent expectation values of the form $\langle c_{i\alpha}(\tau)c_{j\beta}(0)\rangle $. 
We perform these calculations numerically using the EPOCH (Equilibrium Propagator by Orthogonal polynomial CHain) method for computing the equilibrium propagators, i.e.~Green's functions,  directly in the time-domain that we have developed in Ref.~\cite{Method}. In the EPOCH method the equilibrium propagator is expanded in Legendre polynomial mode transients that are recursively computed, allowing us to study both the even- and odd-$\omega$ pair wave amplitudes directly in real space as well as the time domain. As such, this analysis notably goes beyond previous studies of odd-$\omega$ pairing \cite{PhysRevB.76.054522,Eschrig2007,PhysRevB.98.075425,cayao2019odd}, since the treatment is explicitly in the more natural time domain and also fully quantum mechanical.

%%%%%%%%%%%%%%%%%%%%%%%%%%%%%%%
%                                    RESULTS                                        %
%%%%%%%%%%%%%%%%%%%%%%%%%%%%%%%

\section{Pair amplitudes in clean and disordered NS junctions}
\label{sect2}
To ultimately understand the pair wave amplitude behavior in the presence of disorder in NS junctions, we first study its time evolution without disorder, $\delta/t=0$. Then, we analyze the disordered regime by varying  the disorder strength $\delta/t$.
%%%%%%%%%%%%%%%%%%%%%%%%%%%%%%%
%                                CLEAN CASE                                       %
%%%%%%%%%%%%%%%%%%%%%%%%%%%%%%%
\subsection{Clean regime}
% FIGURE:
\begin{figure}[!t]
	\raggedright
	%\centering
	\includegraphics[width=.5\textwidth]{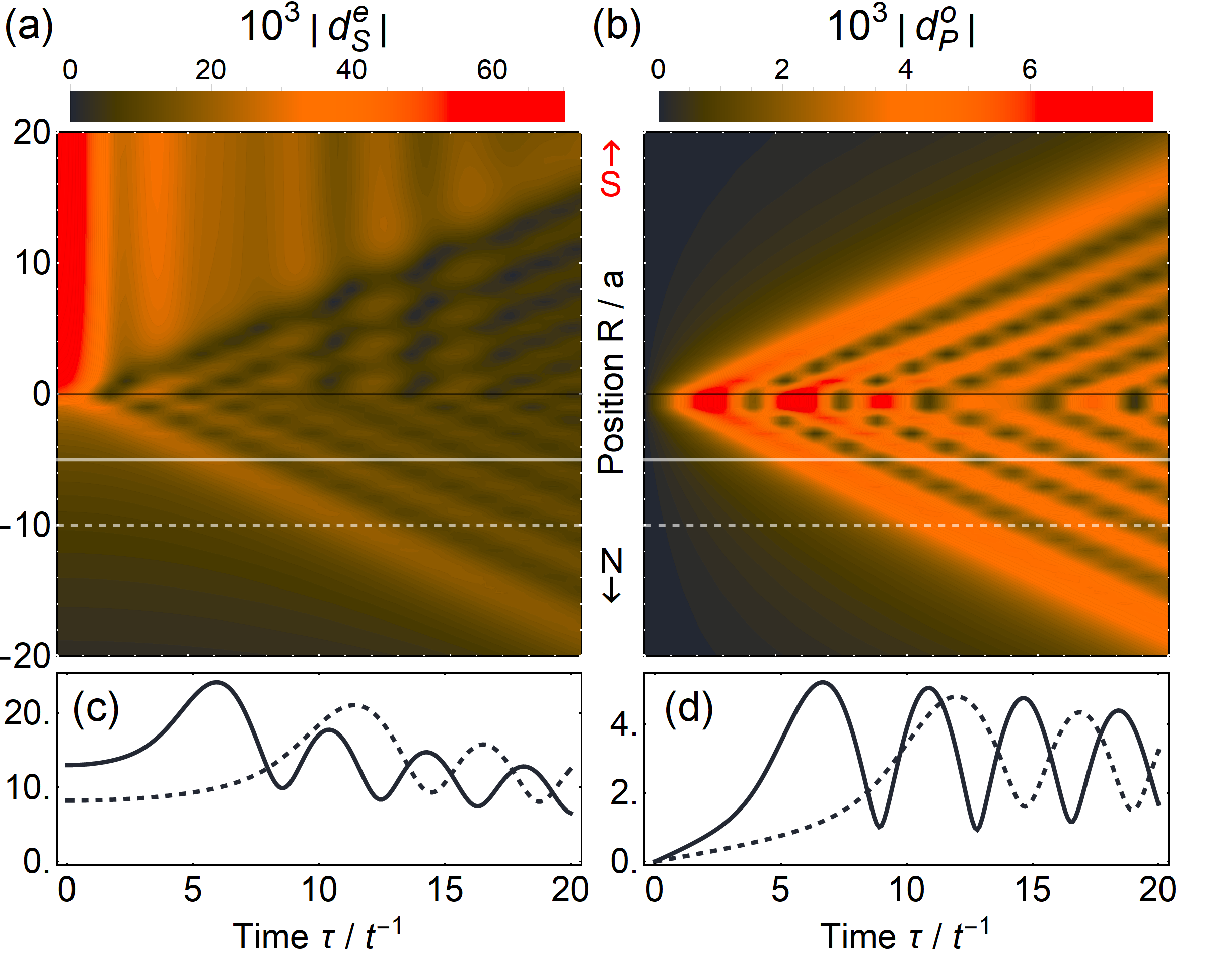}
	\caption{Even-$\omega$ $s$-wave (a)  and odd-$\omega$ $p$-wave (b) pair magnitudes as a function of space and time in clean regime 1D NS junction with interface at $x=0$. (c,d) Time evolution of the even-  and odd-$\omega$ pair magnitudes, respectively, with  solid (dashed) curve corresponding to the position in N indicated by solid (dashed) lines in (a,b). Parameters: $\mu=0.25t$, $L_{\rm N}=250$, $L_{\rm S}=250$, $\Delta=0.01t$. }
	\label{Fig2}
\end{figure}
\begin{figure*}[!t]
	\raggedright
	%\centering
	\includegraphics[width=.995\textwidth]{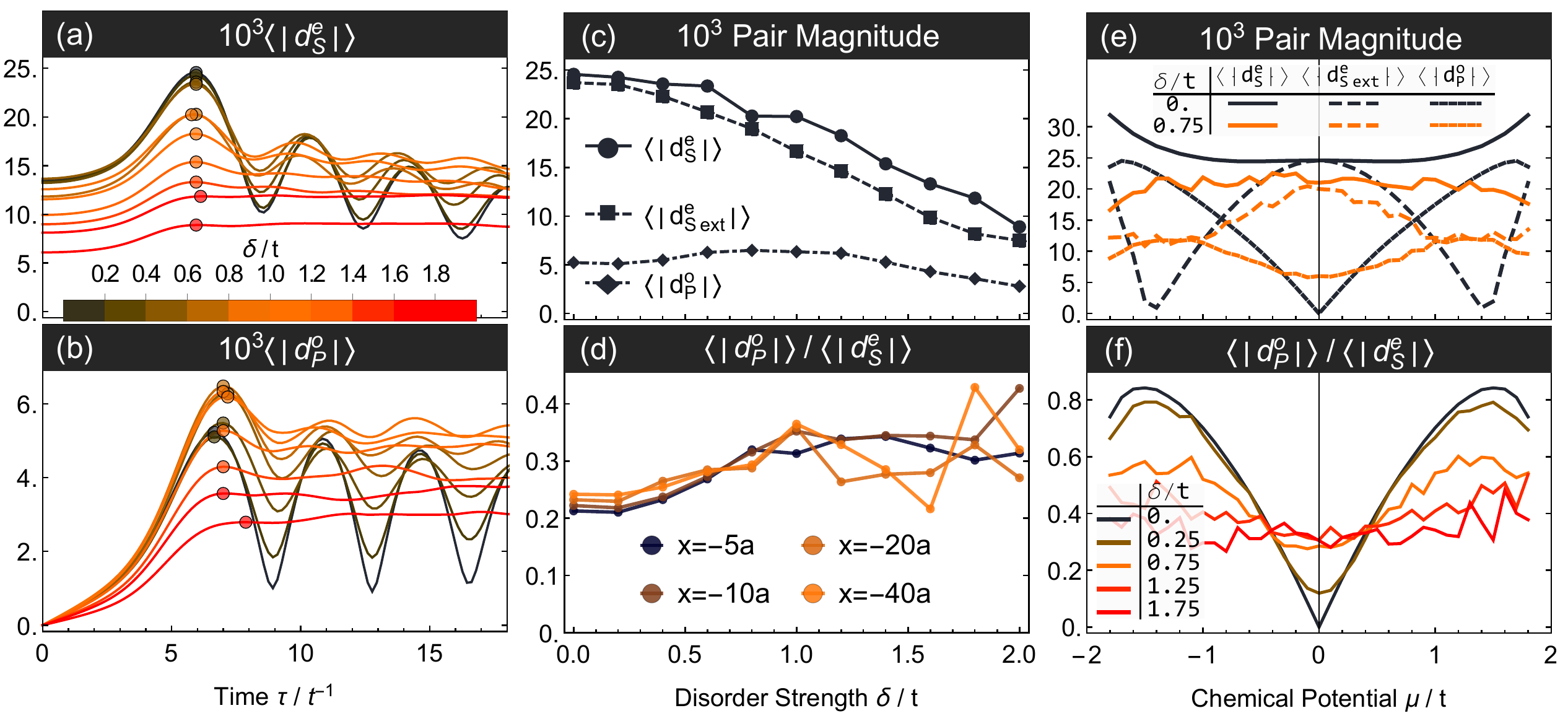}
	\caption{Even- (a) and odd-$\omega$ (b) pair magnitudes at $x =-5a$ in disordered N region in a 1D NS junction, averaged over 300 disorder realizations. Height of the first correlation peak, marked with circles in (a,b), as a function of disorder strength (c) and chemical potential (e) for $s$-, extended-$s$-, and $p$-waves.  Ratio between $p$-wave and $s$-wave magnitudes as a function of disorder strength (d) and chemical potential (f). Unless otherwise stated, same parameters as in Fig.~\ref{Fig2}.
	}
	\label{Fig3}
\end{figure*}
In Fig.\,\ref{Fig2} we show space and time evolution of the magnitude of the even-$\omega$ $s$-wave and odd-$\omega$ $p$-wave pair wave amplitudes, $|d^{e,o}(\tau,R)|$ in a clean 1D NS junction. We explicitly find no other finite amplitudes with $|r| \leq 1$, apart from even-$\omega$ extended $s$-wave correlations, which show similar behavior as the $s$-wave pairing. Notably, the $p$-wave term is not only non-local and odd in space, but also non-local and odd in the relative time coordinate $\tau$, fully consistent with previous results studying the time-dependence of odd-$\omega$ pair amplitudes \cite{PhysRevB.100.144511}.   
At  $\tau=0$ the even-$\omega$ $s$-wave magnitude  exhibits a large and constant value in S, due to the finite order parameter $\Delta$ in S, as see in Fig.\,\ref{Fig2}(a). The odd-$\omega$ pair magnitude in (b), on the other hand, is necessarily zero at $\tau=0$. 
For finite $\tau$ both pair wave amplitudes develop a fan-shaped profile in both S and N regions, albeit with smaller odd-$\omega$ magnitudes. 
The fan profile is due to wave packet propagation in time and space, with a decay as we move away from the NS interface, since the superconducting proximity effect diminishes with distance \cite{Klapwijk2004}. 
There is also an oscillatory pattern, seen as resonances in Fig.\,\ref{Fig2}(c,d), due to interference between incident and reflected quasiparticle states at the NS interface, as noted before \cite{PhysRev.175.559,PhysRevB.92.205424,PhysRevB.98.075425,PhysRevB.100.144511}. In particular, the resonance peaks in the N region are a result of the following time evolution process: an electron at $x_1$ in N has to first propagate towards the NS interface, where it is reflected back as a hole through Andreev reflection, and then propagate back to $x_{2}$ in N in order to contribute to the pair wave amplitude \footnote{Normal reflection is also allowed at the NS interface but can be suppressed in ballistic junctions, and, as shown in Ref.\,\cite{PhysRevB.98.075425}, the coexistence of even- and odd-$\omega$ correlations can be fully understood in terms of Andreev reflection.}. A simple analytic calculation using a 1D continuum model yields the expected time of this process to $\tau_{0} = d/v_{\rm F}$, where  $d =|x_{1}|+|x_{2}|$ is the round trip distance and $v_{\rm F}$ the Fermi velocity, see Supplemental Material (SM)  for a complete derivation \cite{SM}. This value of $\tau_0$ is fully consistent with the results in Fig.\,\ref{Fig2}(c,d). The overall result is that large pair magnitudes disperse linearly away from the interface in both the N and S regions, in agreement with the role of the NS interface as the generator of even- and odd-$\omega$ correlations \cite{PhysRevB.76.054522,Eschrig2007,PhysRevB.98.075425,cayao2019odd}. 

%%%%%%%%%%%%%%%%%%%%%%%%%%%%%%%
%                                DISORDERED CASE                             %
%%%%%%%%%%%%%%%%%%%%%%%%%%%%%%%
\subsection{Disorder regime}
Next we turn to our main topic on how the induced pair correlations are affected by disorder in the N region. Our goal is to ascertain if the odd-$\omega$ $p$-wave correlations are very fragile to disorder as previously assumed \cite{PhysRevLett.98.037003,golubovreview,Nagaosa12}. To this end, we investigate how the odd-$\omega$ $p$-wave correlations are affected by random scalar potential disorder. As a control, we also investigate how both the on-site and extended $s$-wave pair correlations are affected, since both of these isotropic wave states are assumed to be disorder robust. The extended $s$-wave also provides additional control in that it has the same spatial extent as the odd-$\omega$ $p$-wave, both being nearest-neighbor correlations. Our main results are presented in Fig.\,\ref{Fig3}, where we plot the induced pair correlations in a 1D NS system for a wide range of disorder strength $\delta/t$, from the clean limit $\delta/t=0$ to the strongly disordered limit $\delta/t=2$. For any finite $\delta/t$, the disorder produces a finite elastic mean free path $l_e \approx  4.3a (t/\delta)^2$, see SM \cite{SM}. Our range of disorder therefore spans both the regimes where the mean free path is significantly shorter and longer than the superconducting coherence length $\xi = \hbar v_{F} / \Delta \sim 20a$. 		
	
We first present the time evolution of the disorder-averaged pair magnitudes with the on-site even-$\omega$ $s$-wave in Fig.\,\ref{Fig3}(a) and the odd-$\omega$ $p$-wave in Fig.\,\ref{Fig3}(b). Both magnitudes are taken at the fixed spatial position indicated by the solid white line in Fig.\,\ref{Fig2}. At the vanishing disorder strength, we recover the same situation as discussed in Fig.\,\ref{Fig2}, where both the even and odd-$\omega$ pair magnitudes are finite in the N region and exhibit strong oscillations in time. For both magnitudes, the oscillatory patterns in time gradually diminish with increased disorder, but the changes in the magnitudes are clearly different between the even and odd-$\omega$ correlations. Extracting the pair magnitudes at their first resonance peak, indicated by the circle markers in Fig.~\ref{Fig3}(a,b) and plotted in Fig.~\ref{Fig3}(c) as a function of disorder strength, we show that, while the on-site and extended $s$-wave correlations are both monotonically decreasing with the disorder strength, the odd-$\omega$ $p$-wave correlations are non-monotonic with an initial enhancement with increasing disorder. Thus, as we present in Fig.~\ref{Fig3}(d), the ratio of odd-$\omega$ $p$-wave to even-$\omega$ $s$-wave correlations actually {\it increases} with the disorder strength. In particular, the odd to even ratio undergoes a rapid growth around the crossover $l_e \sim \xi$, appearing near $\delta/t \approx 0.5$, and is monotonically increasing for all $l_e/\xi$, all the way to the dirty limit $l_e \ll \xi$. This is in complete contrast to the expected disorder fragility of all anisotropic non-$s$-wave states \cite{PhysRevLett.98.037003,golubovreview,Nagaosa12} and clearly shows that the $p$-wave pairs cannot be ignored even in the dirty limit. The relative increase of the odd-$\omega$ $p$-wave correlations is sustained throughout the N region, even beyond the superconducting coherence length $\xi \sim 20a$, as is clearly evident from the ratios calculated at multiple distances away from the NS junction in Fig.~\ref{Fig3}(d). This rules out a mere interface effect.

To examine the stability of these disorder effects for different parameter values, we present in Fig.\,\ref{Fig3} (e,f) the disordered averaged pair magnitudes as a function of the chemical potential $\mu$, again at the first resonance peak (circles in Fig.\,\ref{Fig3}(a,b)). In the clean regime the even-$\omega$ $s$-wave pair magnitude is almost constant for all values of $\mu$, while the  odd-$\omega$ $p$-wave amplitude identically vanishes at half-filling, $\mu =0$. The latter is due to particle-hole symmetry in the normal state at $\mu =0$, resulting in a cancellation of odd-time correlations \footnote{With particle-hole symmetry in the normal state the process of traveling in N towards the junction as an electron and the process traveling away from the interface as a hole identically cancel in the anomalous Green's function, resulting in zero odd-time pair correlations.}. While this cancellation of the odd-$\omega$ amplitude at $\mu =0$ in the clean limit occurs in a fine-tuned situation, what is highly relevant for the general understanding of $p$-wave pair correlations, is that they still become finite at $\mu =0$ when introducing disorder. Thus disorder can even {\it generate} $p$-wave pairing. Furthermore, as seen in Fig.~\ref{Fig3}(f), where we plot the $p$-wave to $s$-wave ratio, the $p$-wave pair correlations are large, at around 40\%, and essentially constant for disorder ranging from moderately weak to very strong. These results highlight the crucial role of non-magnetic disorder, with disorder not suppressing odd-$\omega$ $p$-wave pairing, but even generating it. Additionally, we have also verified that qualitatively both the clean and disordered behaviors are independent of the size of the superconducting order parameter for all $\Delta < t$

	The exceptional disorder robustness of the odd-$\omega$ $p$-wave correlations is a consequence of the specific mechanism by which they appear. In the clean limit, $p$-wave pair correlations in NS junctions are generated due to the spatial translation symmetry breaking at the interface. Fermion statistics dictates that these $p$-wave correlations must have an odd-$\omega$ symmetry. Introducing non-magnetic disorder in the N region cause two diametrically opposing effects. First, disorder reduces the propagation of Cooper pairs, especially anisotropic pairs. Secondly, disorder further disrupts the translation symmetry and thus generates more odd-$\omega$ $p$-wave correlations. The exact balance between these two effects can only be handled numerically and our results show that the generation even outweighs the suppression in propagation of $p$-wave pairs. Thus odd-$\omega$ spin-singlet $p$-wave pairing is remarkably robust to disorder in dirty NS junctions.  This result is in stark contrast to the dirty limit, or Usadel, formulation of quasiclassical theory, where only isotropic $s$-wave pairing states are explicitly considered, while any anisotropic pairing terms are assumed to vanish in the dirty limit $l_e/\xi \ll 1$ \cite{PhysRevLett.98.037003, golubovreview, Nagaosa12}. It hence follows that any proper quasiclassical treatment has to include anisotropic pair channels using e.g.~the Eilenberger equations \cite{PhysRevB.58.14531,Keser2015}. We also stress that our results are not in contradiction to  Anderson's theorem \cite{PhysRevLett.25.507}, as the pair amplitudes in N are proximity-induced, while Anderson's theorem only formally applies to  bulk superconductors.

%%%%%%%%%%%%%%%%%%%%%%%%%%%%%%%
%                                          2D                                              %
%%%%%%%%%%%%%%%%%%%%%%%%%%%%%%%
\subsection{Beyond 1D}
In the above we focused on the results for 1D NS junctions, but our main findings also hold for NS junctions in higher dimensions. In the SM \cite{SM} we show plots that are equivalent to Fig.~\ref{Fig2}-\ref{Fig3} but for a 2D NS junction and that are analogously obtained by calculating the time evolution of Eq.\,(\ref{Eq:1}). We have likewise verified similar results for 3D junctions. 
	The fact that we find similar results across multiple spatial dimensions means that our main results are independent of any localization effects that can otherwise be pronounced in low dimensional systems \cite{PhysRevLett.42.673}, since the localization lengths are of different orders of magnitude in different dimensions \cite{Kroha1990}.	
	
	\begin{figure}[!t]
		%\raggedright
		\centering
		\includegraphics[width=.45\textwidth]{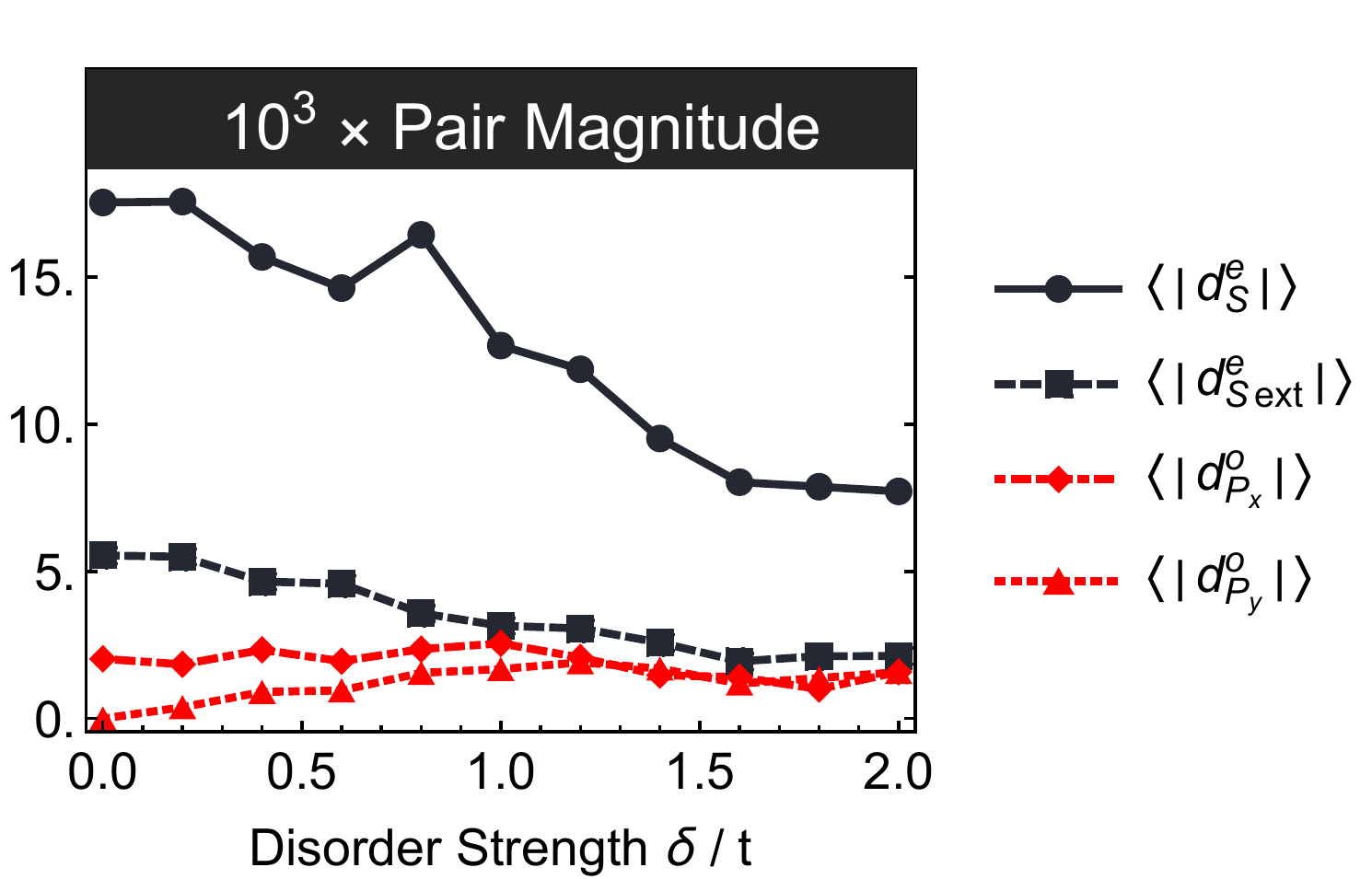}
		\caption{Disorder-averaged $s$-wave (black) and $p$-wave (red) pair magnitudes at $x=-5a$ as a function of disorder strength in a 2D SN junction, taken as the height of the first peak in the time evolution, averaging over 15 disorder realizations. Parameters: $\mu=0.5t$, $L_{\rm N_{x,y}}=50$, $L_{\rm S_{x,y}}=50$, $\Delta=0.01t$. }
		\label{Fig4}
	\end{figure}

While the results in higher dimensions are strikingly similar to those in 1D, and also more stable due to increased self-averaging, there are some importance differences.
	In addition to the on-site and extended even-$\omega$ $s$-waves and the odd-$\omega$ $p_x$-wave correlations previously seen in 1D, there exist in 2D also a odd-$\omega$ $p_y$-wave correlations that is oriented along the junction interface, see Fig.\ref{Fig1}. In Fig.\,\ref{Fig4} we plot the induced disorder-averaged pair magnitudes versus disorder strengths in a 2D SN junction, now also with the $p_y$-wave correlations. The plot can be directly compared to the analogous plot in Fig.\ref{Fig3} for 1D, given the very similar elastic mean free path in the 2D, $l_e \approx  4.7a (t/\delta)^2$, see SM \cite{SM}. In the clean limit $\delta/t=0$, the $p_y$-wave magnitude is  completely absent, because the interface only breaks the translation invariance along the $x$-direction and therefore no correlations with a $p_y$-wave symmetry can be generated. Still, at finite disorder, we see that finite odd-$\omega$ $p_y$-wave pair amplitudes are generated alongside the initial enhancement of the already finite odd-$\omega$ $p_x$-wave amplitude. In contrast, the presumed disorder robust isotropic $s$-wave states instead show a pronounced monotonic suppression with increasing disorder. This demonstrates again that the odd-$\omega$ $p$-wave pair amplitudes are highly disorder robust and always constitute a significant portion of the induced pairing in disordered NS junctions.
	The emergence of odd-$\omega$ correlations along the $y$-axis with disorder can generally be understood as arising from the additional breaking of translational invariance along the $y$-axis by the disorder and its resulting scattering of pair wave amplitudes into the $p_y$-wave channel, although we point out that this alone is not enough. In fact, random charge potential disorder cannot by itself generate odd-$\omega$ pair correlations, as shown in Ref.~\onlinecite{PhysRevB.100.144511}.  What is needed in addition is the NS junction and its associated spatial gradient in the superconducting order parameter, which therefore plays an important role in the disorder generation of the odd-$\omega$ $p_y$-wave pair amplitudes.

%%%%%%%%%%%%%%%%%%%%%%%%%%%%%%%%%%%
%                                         CONCLUSIONS                                      %
%%%%%%%%%%%%%%%%%%%%%%%%%%%%%%%%%%%
\section{Conclusions}%
\label{conclusions}
In conclusion, we have demonstrated that odd-$\omega$ spin-singlet $p$-wave pair correlations are robust against random scalar, non-magnetic, disorder in conventional and dirty NS junctions. In fact, the $p$-wave pair correlations display a non-monotonic dependence on the disorder strength, with an initial enhancement and also constitute a growing fraction of the induced pair correlations in the N region, notably even in the highly disordered regime where the elastic mean free path is much smaller than the superconducting coherence length. The $p$-wave pair correlations therefore display an even greater disorder robustness than the isotropic $s$-wave components that are instead monotonically decreasing with an increased disorder. As a consequence, odd-$\omega$ $p$-wave pair correlations always constitute a significant portion of the proximity-induced superconductivity in dirty NS junctions. With properties, such as the Josephson and Meissner effects, being directly dependent on the superconducting pair amplitudes, we expect our results to have interesting consequences for various NS systems and superconducting hybrid structures.

%%%%%%%%%%%%%%%%%%%%%%%%%%%%%%%%%%%
%                             ACKNOWLEDGEMENTS                                    %
%%%%%%%%%%%%%%%%%%%%%%%%%%%%%%%%%%%

\section{Acknowledgements}
We thank A.~Balatsky, D.~Chakraborty, and T.~L\"{o}fwander for useful discussions. We acknowledge financial support from the Swedish Research Council (Vetenskapsr\aa det Grant No.~2018-03488), the European Research Council (ERC) under the European Unions Horizon 2020 research and innovation programme (ERC-2017-StG-757553), the Knut and Alice Wallenberg Foundation through the Wallenberg Academy Fellows program and the EU-COST Action CA-16218 Nanocohybri.

\bibliography{biblio}
%\cleardoublepage
\onecolumngrid
\appendix



\section*{Supplementary material}

# Supplemental Material for "Disorder-robust $p$-wave pairing with odd frequency dependence in normal metal-conventional superconductor junctions"

Tomas Löthman,[1] Christopher Triola,[2] Jorge Cayao,[1, *] and Annica M. Black-Schaffer[1]

[1]*Department of Physics and Astronomy, Uppsala University, Box 516, S-751 20 Uppsala, Sweden*
[2]*Los Alamos National Laboratory, Los Alamos, New Mexico 87544, USA*
(Dated: July 11, 2021)

In this supplemental material we provide details to further support the results and conclusions of the main text.

## I. ANALYTIC TREATMENT OF PAIR AMPLITUDES IN CLEAN NORMAL METAL-SUPERCONDUCTOR JUNCTIONS

In the main text we presented numerical results for the proximity-induced superconducting pair amplitudes in the normal region of normal metal-superconductor (NS) junctions, modeled using a tight-binding Hamiltonian. An important advantage of our numerical approach is that it allows us to easily consider both the clean and disordered regimes, with no approximations. However, since numerical analyses can only be performed for a finite number of parameters, insight can often be gained from a complementary approximate analytic calculation. In this Section we study the induced pair amplitudes in a one-dimensional (1D) NS junction in the clean regime analytically using perturbation theory.

### A. Bare Green's functions

To facilitate our analytic treatment we assume that both sides of the NS junction are infinite and well described by continuum models with dispersions given by $\xi_{\mathbf{k}} = k^2/2m - \mu$ and $E_{\mathbf{k}} = \sqrt{\xi_{\mathbf{k}}^2 + \Delta^2}$, with $\hbar = 1$, for the N and S regions, respectively. Here, $m$ denotes the effective mass of the electrons in both the N and S regions, $k$ is the wave vector, $\mu$ is the chemical potential, and $\Delta$ is the superconducting order parameter in the S region. Given these dispersions, and first assuming no tunneling between the two regions, it is straightforward to show that the Nambu space Green's functions in the N and S regions are given by[1, 2]:

$$
\begin{aligned}
\hat{\mathcal{G}}_S^{(0)}(\mathbf{k}; z) &= \frac{z\hat{\tau}_0 + \left(\frac{k^2}{2m} - \mu\right)\hat{\tau}_3 + \Delta\hat{\tau}_1}{z^2 - \left(\frac{k^2}{2m} - \mu\right)^2 - \Delta^2}, \\
\hat{\mathcal{G}}_N^{(0)}(\mathbf{k}; z) &= \frac{z\hat{\tau}_0 + \left(\frac{k^2}{2m} - \mu\right)\hat{\tau}_3}{z^2 - \left(\frac{k^2}{2m} - \mu\right)^2},
\end{aligned}
\tag{1}
$$

where $\hat{\tau}_i$ ($\hat{\tau}_0$) is the $i^{\text{th}}$ Pauli matrix (identity matrix) in the 2×2 particle-hole space and $z$ is a complex frequency allowing us to keep track of Matsubara ($z = i\omega_n$), retarded ($z = \omega + i0^+$), and advanced ($z = \omega - i0^-$) Green's functions, simultaneously. Note that the pair amplitude, $F$, is simply given by the off-diagonal matrix element of the Nambu space Green's function, and, therefore, it is trivially zero in the N region and given by the term proportional to $\hat{\tau}_1$ in the S region in the limit of no tunneling between the two regions.

### B. NS junction

To study proximity-induced pairing in the N region, we assume the NS interface allows local quasiparticle tunneling between the two systems at $x = 0$. In this case, the Dyson equation for the full Nambu-space Green's function in the N region may be written in real space as:

$$
\hat{\mathcal{G}}_N(x_1, x_2; z) = \hat{\mathcal{G}}_N^{(0)}(x_1 - x_2; z) + t_0^2 \hat{\mathcal{G}}_N^{(0)}(x_1; z)\hat{\tau}_3 \hat{\mathcal{G}}_S^{(0)}(0; z)\hat{\tau}_3 \hat{\mathcal{G}}_N(0, x_2; z).
\tag{2}
$$

---

* Corresponding author: jorge.cayao@physics.uu.se

This equation can be iterated to obtain a power series expression for $\hat{\mathcal{G}}_N(x_1, x_2; z)$ to arbitrary order in the tunneling, $t_0$. Since we are only interested in the qualitative behavior of these expressions we truncate this expansion to leading order in $t_0$:

$$\hat{\mathcal{G}}_N(x_1, x_2; z) \approx \hat{\mathcal{G}}_N^{(0)}(x_1 - x_2; z) + t_0^2 \hat{\mathcal{G}}_N^{(0)}(x_1; z)\hat{\tau}_3 \hat{\mathcal{G}}_S^{(0)}(0; z)\hat{\tau}_3 \hat{\mathcal{G}}_N^{(0)}(x_2; z). \tag{3}$$

The bare Green's functions on the right-hand side of Eq. (3) may be obtained by Fourier transforming the $k$-space expressions in Eqs. (1) to real space giving:

$$\hat{\mathcal{G}}_S^{(0)}(0; z) = -\frac{im \, \mathrm{sgn}\,(\mathrm{Im}\Omega)}{2k_F \Omega} \left[ z g_+^S(z)\hat{\tau}_0 + \Delta g_+^S(z)\hat{\tau}_1 + \Omega g_-^S(z)\hat{\tau}_3 \right],$$
$$\hat{\mathcal{G}}_N^{(0)}(x; z) = -\frac{im}{2k_F} \left[ g_+^N(x; z)\hat{\tau}_0 + g_-^N(x; z)\hat{\tau}_3 \right], \tag{4}$$

where

$$g_\pm^S(z) = \frac{1}{\sqrt{1 + \frac{\Omega}{\mu}}} \pm \frac{1}{\sqrt{1 - \frac{\Omega}{\mu}}},$$
$$g_\pm^N(x; z) = \frac{e^{i\mathrm{sgn}(\mathrm{Im}z)|x|k_F\sqrt{1 + \frac{z}{\mu}}}}{\sqrt{1 + \frac{z}{\mu}}} \pm \frac{e^{-i\mathrm{sgn}(\mathrm{Im}z)|x|k_F\sqrt{1 - \frac{z}{\mu}}}}{\sqrt{1 - \frac{z}{\mu}}}, \tag{5}$$
$$\Omega = \sqrt{z^2 - \Delta^2},$$
$$k_F = \sqrt{2m\mu}.$$

These expressions can now be inserted into Eq. (3) to find analytic expressions for the Nambu space Green's function in the N region. Note that the bare Green's functions in real space given by Eqs. (4) depend solely on one coordinate as they are the Fourier transform of Eqs. (1) which describe continuum (infinite) systems where translational invariance is not broken.

### C. Proximity-induced pair amplitudes

As we are primarily interested in the induced pair amplitudes in the N region, we focus on the anomalous component of the Green's function in Eq. (3), which is given by:

$$F_N(x_1, x_2; z) = -\frac{it_0^2 \Delta}{2v_F^3 \Omega} \frac{\mathrm{sgn}\,(\mathrm{Im}\Omega) \exp\left[ i\mathrm{sgn}(\mathrm{Im}z)\left( |x_1|k_F\sqrt{1 + \frac{z}{\mu}} - |x_2|k_F\sqrt{1 - \frac{z}{\mu}} \right) \right]}{\sqrt{1 + \frac{z}{\mu}}\sqrt{1 - \frac{z}{\mu}}} \\ \times \left( \frac{1}{\sqrt{1 + \frac{\Omega}{\mu}}} + \frac{1}{\sqrt{1 - \frac{\Omega}{\mu}}} \right), \tag{6}$$

where $v_F = k_F/m$. This expression for the proximity-induced pairing in the N region is exact to order $t_0^2$. However, we would like to simplify this expression to gain a better understanding of the physical properties of these pair amplitudes. To proceed, we note that the pair amplitudes are strongly peaked for frequencies $z \sim \Delta$. Moreover, we recall that, in typical superconductors, $\Delta << \mu$. Therefore, we assume that $z/\mu, \Delta/\mu << 1$ and expand all terms to leading order in these quantities. After a fair amount of algebra we obtain the approximate expression:

$$F_N(x_1, x_2; z) \approx -\frac{it_0^2 \Delta}{v_F^3 \Omega} \mathrm{sgn}\,(\mathrm{Im}\Omega) \, e^{i\mathrm{sgn}(\mathrm{Im}z)z\frac{|x_1| + |x_2|}{v_F}} \{ \cos\left[\mathrm{sgn}(\mathrm{Im}z)k_F\left(|x_1| - |x_2|\right)\right] \\ + i\sin\left[\mathrm{sgn}(\mathrm{Im}z)k_F\left(|x_1| - |x_2|\right)\right] \}, \tag{7}$$

which is accurate to leading order in $t_0$, $z/\mu$, and $\Delta/\mu$.

### D. Proximity-induced pair amplitudes in the time domain

Now that we have a relatively compact expression for the proximity-induced pairing in the N region, we wish to make contact with the numerical results in the main text by Fourier-transforming to the time domain. First, we recall that, in general, a time-ordered propagator is related to the retarded and advanced propagators by:

$$g^T(t) = \int_{-\infty}^{\infty} \frac{d\omega}{2\pi} e^{-i\omega t} \left[ \theta(\omega) g^R(\omega) + \theta(-\omega) g^A(\omega) \right]. \tag{8}$$

Using this relation, together with the expression in Eq. (7) the time-ordered pair amplitude in the N region may be written as:

$$F_N^T(x_1, x_2; t) = -\frac{it_0^2 \Delta}{v_F^3} \int_0^{\infty} \frac{d\omega}{2\pi} e^{i\omega \tau_0} \left\{ \frac{e^{-i\omega t} e^{ik_F(|x_1| - |x_2|)} + e^{i\omega t} e^{-ik_F(|x_1| - |x_2|)}}{\sqrt{(\omega + i0^+)^2 - \Delta^2}} \right\}, \tag{9}$$

where we define $\tau_0 = \frac{|x_1| + |x_2|}{v_F}$, which corresponds to the total time for a quasiparticle at position $x_1$ in N to travel towards the NS interface and then back to position $x_2$ in N, as discussed in the main text. By inspection of the terms within the curly braces in Eq. (9), we see that the proximity-induced pair amplitude possesses both even and odd components in the time variable, $t$. These even- and odd-time components are given by:

$$F^e(x_1, x_2; t) = -\frac{i2t_0^2 \Delta}{v_F^3} \cos \left[ k_F \left( |x_1| - |x_2| \right) \right] \int_0^{\infty} \frac{d\omega}{2\pi} \frac{e^{i\omega \tau_0} \cos(\omega t)}{\sqrt{(\omega + i0^0)^2 - \Delta^2}},$$
$$F^o(x_1, x_2; t) = -\frac{i2t_0^2 \Delta}{v_F^3} \sin \left[ k_F \left( |x_1| - |x_2| \right) \right] \int_0^{\infty} \frac{d\omega}{2\pi} \frac{e^{i\omega \tau_0} \sin(\omega t)}{\sqrt{(\omega + i0^+)^2 - \Delta^2}}. \tag{10}$$

Focusing on the odd-time/odd-frequency component, and changing integration variables to $u = \omega/\Delta$, we find:

$$\begin{aligned} F^o(x_1, x_2; t) &= \frac{-t_0^2 \Delta}{2\pi v_F^3} \sin \left[ k_F \left( |x_1| - |x_2| \right) \right] \left\{ \int_0^{\infty} du \frac{e^{iu\Delta(\tau_0 + t)} - e^{iu\Delta(\tau_0 - t)}}{\sqrt{(u + i\eta)^2 - 1}} \right\} \\ &\approx \frac{-t_0^2 \Delta}{2\pi v_F^3} \sin \left[ k_F \left( |x_1| - |x_2| \right) \right] \left\{ -i \int_0^1 du \left[ e^{iu\Delta(\tau_0 + t)} - e^{iu\Delta(\tau_0 - t)} \right] \right. \\ &\left. + \int_1^{\infty} du \frac{e^{iu\Delta(\tau_0 + t)} - e^{iu\Delta(\tau_0 - t)}}{u} \right\}, \end{aligned} \tag{11}$$

where in the second line we have written the exact integral on the interval $[0, \infty)$ as two approximate integrals: (i) the interval where we assume $\omega/\Delta \leq 1$, and (ii) the interval where we assume $\omega/\Delta >> 1$. Simplifying these expressions we can write the odd-time/odd-frequency pair amplitude in the N region as:

$$\begin{aligned} F^o(x_1, x_2; t) &= \frac{-t_0^2 \Delta}{2\pi v_F^3} \sin \left[ k_F \left( |x_1| - |x_2| \right) \right] \left\{ E_1 \left[ -i\Delta(\tau_0 + t) \right] - E_1 \left[ -i\Delta(\tau_0 - t) \right] \right. \\ &\left. + 2 \frac{e^{i\Delta \tau_0} \left[ t \cos(t\Delta) - i\tau_0 \sin(t\Delta) \right] - t}{\Delta(\tau_0^2 - t^2)} \right\}, \end{aligned} \tag{12}$$

where $E_1(z) = \int_1^{\infty} e^{-tz}/t dt$ is the exponential integral.

One of the most important features to observe from Eq. (12) is that the last term gives rise to a sharp peak in time, at $t = \tau_0 = \frac{|x_1| + |x_2|}{v_F}$, in full agreement with the numerical results presented in the main text. In fact, this peak has an intuitive origin in the physical processes giving rise to all pair amplitudes in the N region: $\tau_0$ is the amount of time it takes a quasiparticle to complete a trip from $x_1$ to the interface, reflect as a hole at the interface, and travel back to the point $x_2$. Additionally, we note that the expressions given by Eqs. (10) for the even- and odd-time pair amplitudes reveal an oscillatory behavior in both time and space, consistent with the numerical results presented in the main text for the clean regime in 1D. Furthermore, we see here that the periodicity of the spatial oscillations is determined by the Fermi wave vector $k_F$, while the periodicity of the temporal oscillations is determined by $\Delta$. Lastly, we point out that a similar analysis can be carried out for higher dimensions, where the main features discussed in this part are expected to hold.

### E. Comparison to numerical results in 1D

In Fig. 2(a,b) in the main text we showed that in the clean regime both the even- and odd-time pair amplitudes develop resonance peaks for a finite time separation, which then propagate away from the NS interface, resulting in a fan-like profile. In this Section we show that the time separation at which the first resonance peak develops, defining the outer structure of the fan, is the travel time for making the round trip distance to the junction at the Fermi velocity of the normal state.

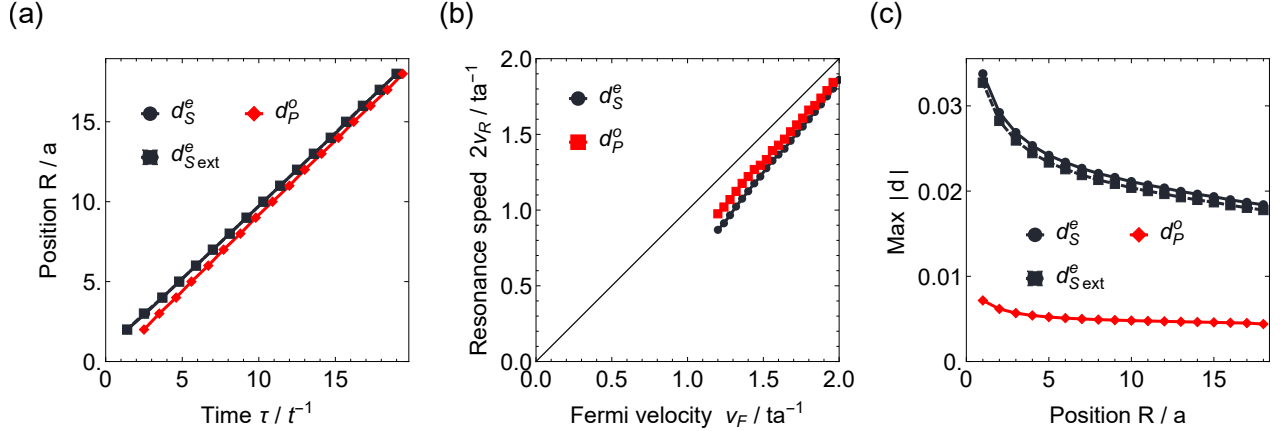

FIG. S1. First resonance peaks of a clean 1D NS junction for the even-time $s$-wave and extended $s$-wave (black) and the odd-time $p_x$-wave (red) pair magnitudes. Parameters are the same as in Fig. 2 in the main text. (a) Resonance peak position as a function of time separation $\tau$. (b) Twice the velocity of the resonance peak $v_R(\mu)$, defined as twice the slope of the line in (a), plotted as a function of the Fermi velocity in N, $v_F(\mu)$, changed by tuning the chemical potential $\mu$. (c) Pair magnitude at the resonance peak as a function of the distance from the NS interface $R$.

We first show in Fig. S1 (a) that with an increasing time separation, the resonance peaks of Fig. 2 in the main text move linearly away from the junction, implying that the resonance peaks have a definite speed, here called $v_R$. According the the analytical calculations presented in the previous section, this speed is the Fermi velocity $v_F$. To test numerically if $v_F$ is indeed the speed in the numerical calculations, it does not suffice to just compute the time separation and position of the resonance peaks as in Fig. S1(a), but rather we need to also manipulate the Fermi velocity $v_F$, because only then can we observe if the resonance propagation of the resonances is dependent on the Fermi velocity.

The Fermi velocity $v_F$ is defined as the slope of the band dispersion, $v_F = |d\epsilon(k)/dk|$ at the Fermi energy. By changing the chemical potential $\mu$ the filling of the band is changed and thus the Fermi energy intersects the energy band $\epsilon(k)$ at new point, with a different slope. Thus, the Fermi velocity is manipulated by the chemical potential $\mu$. If the N region had been infinite, it would have had the band dispersion $\epsilon(k) = 2\cos(k)$ with Fermi velocity $v_F = |d\epsilon(k)/dk| = 2|\sin(k)|$. Even though the N region is finite in the numerical model, we still approach our problem assuming that these ideal expression approximates the band dispersion. With this assumption the Fermi velocity is related to the chemical potential $\mu$ through $v_F = \sqrt{4 - \mu^2}$. We can then numerical calculate the speeds of the resonance peaks $v_R(\mu)$ as the slope of the line in Fig. S1(a) by changing $\mu$. Recognizing that $R/a$ on the $y$-axis in Fig. S1(a) is half the round-trip distance, we in Fig. S1(b) show how the ideal $v_F(\mu)$ relates to $2v_R(\mu)$. We see that the agreement is almost perfectly $v_F(\mu) = 2v_R(\mu)$, exactly as suggested by our analytical results. The minor deviation is easily caused by the fact that changing $\mu$ not only changes $v_F$ but also the low-energy density of states and thus the overall size of the pair amplitudes. To conclude, it is the Fermi velocity that sets the speed of the resonance peaks at the NS junction, and the time-separation for the peaks is given by the round-trip distance $2R$. In closing, we also show in Fig. S1(c) that height of the resonance peaks, i.e. the pair magnitude, slowly decays with distance away from the junction for both odd-$\omega$ and even-$\omega$ due to the inevitable decay of the proximity effect away from the interface.

## II. ADDITIONAL RESULTS FOR 2D NS JUNCTIONS

In the main text we mainly presented the paring amplitudes for 1D NS junctions. Here we show that the main conclusions that we draw from the 1D system also hold in higher dimensions.

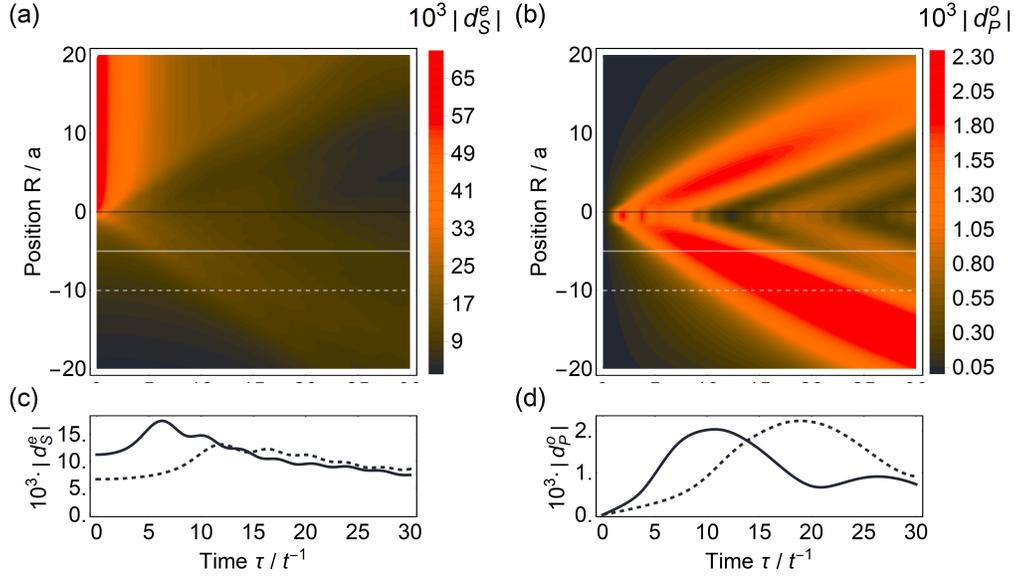

FIG. S2. Even-$\omega$ $s$-wave (a) and odd-$\omega$ $p_x$-wave (b) pair magnitudes in the N region as a function of position from the NS interface $R$ and time $\tau$ for a clean 2D SN interface located at $R = 0$. (c,d) Time evolution of the even- and odd-$\omega$ pair magnitudes, respectively, with solid (dashed) curve corresponding to the fixed position in N indicated by solid (dashed) lines in (a,b). Parameters: $\mu = 0.5t$ and $\Delta = 0.01t$ on a square NS junction with side lengths $L_S = L_N = 100a$.

Starting with the clean regime of a 2D junction, we show, in complete analogy with Fig. 2 in the main text, the propagation of even-$\omega$ $s$-wave and the odd-$\omega$ $p_x$-wave pair magnitudes $|d^{e,o}(\tau, R)|$ in Fig. S1. As for the 1D case, the pair magnitudes are obtained from Eq. (2) in the main text. As seen, most aspects of the propagation remain unchanged between 1D and 2D junctions. At equal times $\tau = 0$, the even-$\omega$ magnitudes in Fig. S2(a) have a largely constant amplitude inside the S region, which spreads into the N region by proximity effect. In contrast, the odd-$\omega$ magnitude at $\tau = 0$ is identically zero by symmetry. At finite time separations $\tau > 0$, we find a ridge of large pair magnitudes, a resonance, linearly emanating away from the NS interface, also in agreement with the 1D results. Thus the broad features remain unchanged between 1D and 2D, although there are also some differences. Most notable from Fig. S2 is the absence of the fan-like oscillatory pattern seen in 1D, and, in addition, the pair resonances are much broader, as clearly seen in the position cuts presented in Fig. S2(c,d). Both of these qualitative differences can be directly interpreted as resulting from the change in dimension. To understand this we return to the physical picture suggested by the pair amplitude time evolution: a quasiparticle making a round-trip from the position $R$ in the N region to the NS interface and then back to $R$. In the 1D junction, there is only one trajectory for this trip. However, in 2D there are a plethora trajectories besides the shortest path, which is the straight line joining the $R$ point and the point on the NS interface closest to it. These 2D trajectories include loops and braids that take full advantage of the extra dimension. However, the total length of these trajectories are always longer than the shortest path and thus their associated travel times will be longer. Still, among the likely paths, most will arrive shortly after the round-trip time for the shortest path. As a consequence, similarly as in the 1D junction, the first resonance peak reflects the travel time of the shortest path. The added variations in travel times in 2D however does two things; first, it considerably broadens the first resonance peak; second, it washes out the sharp fan-like oscillatory pattern seen in 1D. This explains all qualitative differences seen in Fig. S2 compared to the 1D junction. As this picture suggests, we also see in Fig. S2 that the width of the first resonance peak widens with $R$, reflecting that longer paths have more opportunities to deviate. Less evident from Fig. S2, but still worth mentioning, is that in 2D the height of the pair amplitude at the resonance peaks decay faster than in 1D, partly reflecting that the peak is broadened, but also that the pair amplitude wave disperses geometrically in 2D unlike in 1D.

Finally we also compare the disordered regimes for 1D and 2D NS junctions. In Fig. S3 we show how the disorder-averaged pair magnitudes $|\langle d_{s(p)}^{e(o)} \rangle|$ evolve with disorder strength $\delta$, similarly as in Fig. 3 in the main text. The main difference is that in Fig. S3 we show all wave projections, including the even-$\omega$ $s$-wave (a) and extended $s$-wave (b) states, as well as the odd-$\omega$ $p_x$-wave (c) and also the odd-$\omega$ $p_y$-wave (d) states. We see in agreement with the 1D results in the main text that the even-$\omega$ waves are monotonically decreasing (apart from statistical fluctuation) with increasing disorder, while the odd-$\omega$ $p_x$-wave instead shows an initial increase. Additionally, we find that the odd-$\omega$ $p_y$-wave, which by symmetry is identically zero in the clean region, is quickly generated by disorder. Thus, considering

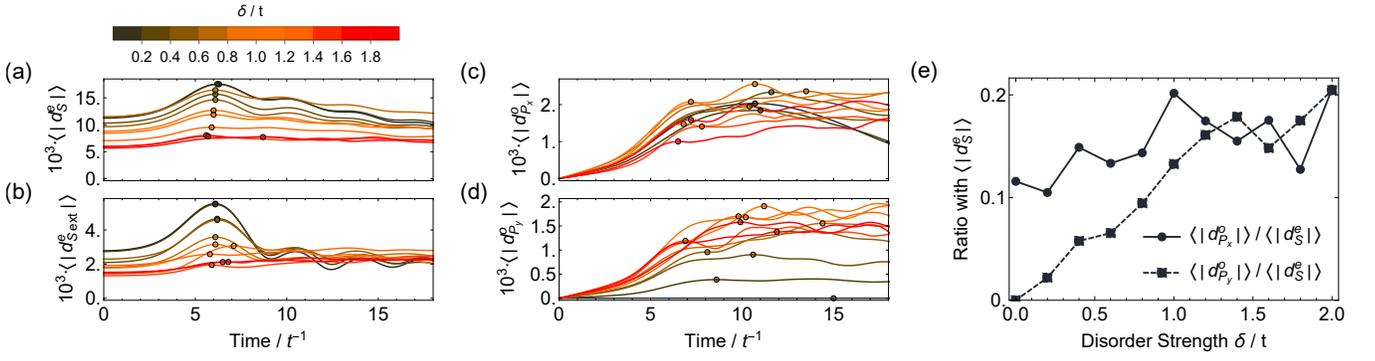

FIG. S3. (a-d) Disorder-averaged pair magnitudes $|\langle d^{e(o)}_{s(p_i)}\rangle|$ at $x = -5a$ in the disordered N region of a 2D NS junction for 15 disorder realizations. (e) The ratio of the first resonance peaks marked by circles in (a-d) for the odd-$\omega$ $p_x$- and $p_y$-wave magnitudes to the even-$\omega$ $s$-wave magnitude. Unless otherwise stated, same parameters as in Fig. 2.

the relative abundance of odd-$\omega$ $p$-wave to the even-$\omega$ $s$-wave pair magnitudes in Fig. S3(e), we see that the ratio of odd- to even-$\omega$ pairing rapidly increases with disorder, with the $p_y$ ratio quickly approaching that of the original odd-$\omega$ $p_x$-wave wave.

## III. ELASTIC MEAN FREE PATH

In this Section we discuss and numerically calculate the elastic mean free path in the disordered normal metal side (N) of the SN junction. Without the superconductor but with a finite disorder strength $\delta/t$, the zero temperature conductivity of the N side of the junction, is given within the Kubo-Greenwood formalism by

$$\sigma(E_F) = e^2 \rho(E_F) l_e(E_F) v_F(E_F) \tag{13}$$

where $l_e(E_F)$ is the elastic mean free path, $v_F(E_F)$ is the Fermi velocity, and $\rho(E_F)$ is the density of states, all measured at the Fermi energy $E_F$. Following Refs. [3, 4], the transport properties and $l_e(E_F)$ are efficiently computed in a fully quantum mechanical manner from the temporal dynamics of a wave-packet by its mean square displacement $\langle \Delta \hat{X}^2(E_F, \tau) \rangle = \langle \delta(E_F - H)[\hat{X}(\tau) - \hat{X}(0)]^2 \rangle / \rho(E_F)$. This is the case because of the way the wave packet spreads in space is directly related to the diffusion coefficient,

$$D(E_F, \tau) = \langle \Delta \hat{X}^2(E_F, \tau) \rangle / \tau \tag{14}$$

which is in turn related to the conductivity through the time-dependent Einstein relation,

$$\sigma(E_F, \tau) = e^2 \rho(E_F) D(E_F, \tau). \tag{15}$$

Therefore, by calculating the diffusion coefficient $D$, it is possible to obtain the elastic mean free path $l_e(E_F)$.

In Fig. S4(a) we present the diffusion coefficient $D(E_F, \tau)$ computed from the time evolution of a stochastic wave-packet on a $L = 10^5 a$ long one dimensional (1D) lattice according to the methods and algorithms of Refs. [3, 4] using 2500 polynomial moments. In a disorder free system, the conduction is ballistic and $D(E_F, \tau) = v_F^2(E_F)\tau$ increases linearly with the elapsed time $\tau$. In contrast, the disorder present in Fig. S4(a) triggers a diffusive regime where, after a short initial time ballistic motion, the diffusion coefficient $D(E_F, \tau)$ saturates due to scattering. From the initial short-time ballistic motion, we calculate the Fermi velocity $v_F(E_F)$ by a least squares fit, and from the saturated maximum value of the diffusion coefficient we extract the elastic mean free path as

$$l_e(E_F) = D_{\max}(E_F)/2v_F(E_F). \tag{16}$$

To get further insights into the behavior of the mean free path, we plot in Fig. S4(b) the elastic mean free path obtained from Eq. (16) as a function of different disorder strength $\delta/t$. As we can see, $l_e(E_F)$ exhibits a monotonically decreasing behavior. The dashed line shows a fit to the inverse square of the disorder strength: $l_e(E_F) = \alpha(t/\delta)^2$. This functional dependence is based on the scattering rate from the Fermi golden rule [5–7] and yields a very good fit for $\alpha \approx 4.3a$. The same relationship also holds for a 2D lattice but there we find $\alpha \approx 4.7a$ using $E_F = 0.5t$. A significant difference between the 1D and 2D cases is, however, the duration over which $D(E, \tau)$ is diffusive before

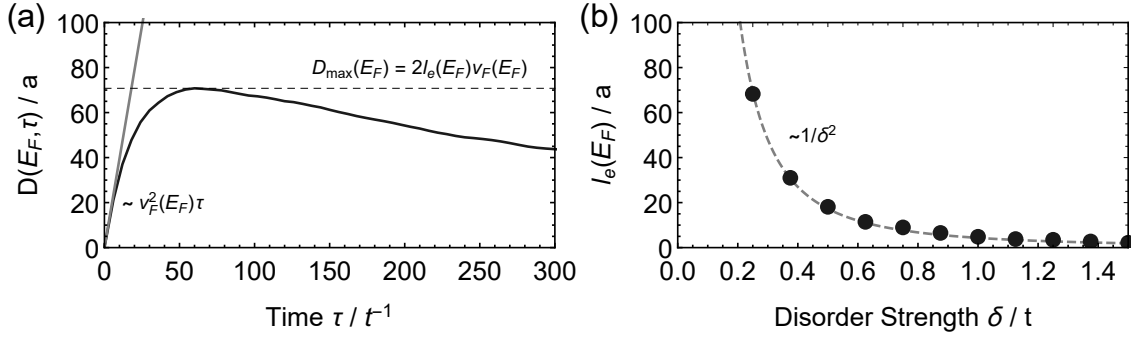

FIG. S4. (a) Diffusion coefficient $D(E_F, \tau)$ as a function of the elapsed time $\tau$ computed from the mean square displacement of a wave packet on a disordered 1D lattice with $10^5$ sites, $\delta = 0.5t$, and $E_F = 0.25t$. There initially exists a short-time ballistic motion where the coefficient $D(E_F) = v_F^2(E_F)t$ grows linearly with the Fermi velocity $v_F$ (solid straight line) after which it saturates. From the saturated value (dashed line) the elastic mean free path can be extracted as $l_e(E_F) = D_{\max}(E_F)/2v_F(E_F)$. (b) Elastic mean free path $l_e(E_F)$ as a function of the disorder strength $\delta/t$ for the same system as in (a). Dashed line shows a fit to the inverse square of the disorder strength: $l_e(E_F) = \alpha(t/\delta)^2$ with $\alpha \approx 4.3a$.

entering a localized regime with its telltale decay from $D_{\max}(E_F)$. This occurs when the mean square displacement saturates,

$$\lim_{\tau \to \infty} \sqrt{\langle \Delta \hat{X}^2(E_F, \tau) \rangle} = \pi \lambda(E_F), \tag{17}$$

which defines the localization length $\lambda(E_F)$ [4, 8]. We find an order of magnitude longer localization lengths $\lambda(E_F)$ in 2D compared with 1D, consistent with earlier results [9], but despite the very similar elastic mean free paths. We use the mean free paths obtained and discussed here to compare with the superconducting coherence length in the main text, Section III B.

---



\end{document}